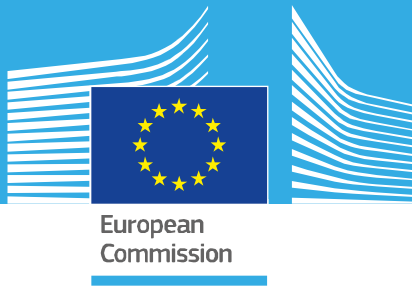

# JRC TECHNICAL REPORT

*Putting Science into Standards (PSIS)*

# Standardisation needs for improved management of COVID-19

*A JRC scoping exercise on potential standardisation gaps*

Jenet, A., Nik, S., Mian L., Schmidtler, S.Z., Annunziato, A, Marin-Ferrer, M., McCourt, J., Lequarre, A.S., Ganesh, A., Taucer, F.

2021

# Contents

## Foreword

The Putting-Science-Into-Standards (PSIS) Workshops is a joint initiative between the European Commission and the European Standardisation Bodies CEN and CENELEC bringing together the scientific, industrial, policy and standardisation communities on a regular basis. The initiative aims to facilitate the identification of emerging science and technology areas which could benefit from standardisation activities to enable innovation and promote industrial competitiveness.

Planning at an early stage and incorporating standardisation can be crucial for accelerating the market uptake of research findings. It is against this background that the European Commission's Joint Research Centre (JRC) has launched a scoping exercise within the JRC, CEN and CENELEC to identity research areas relevant in responding to the ongoing COVID-19 pandemic where standards and harmonisation may play an important role in possible future pandemics.

## Acknowledgements

Authors are grateful to Angel Alvarez for his contribution in graphic design. We wish to thank the Connected and EU-Survey teams for making the JRC-wide call for contributions for the scoping study possible. We also would like to express our gratitude to the JRC in-house Corona Task Force for having disseminated our pledge for contributions.

**Abstract**

The Joint Research Centre (JRC) of the European Commission in collaboration with the European standardisation bodies CEN and CENELEC launched a scoping exercise on standardisation needs in response to COVID-19 and future pandemics.

The purpose of the exercise was to identify ongoing harmonisation initiatives, as was well as further standardisation needs in relevant sectors such as artisanal reusable face masks (CWA 17553), medical face masks, and social distancing in closed public or commercial spaces.

An overview of already ongoing standardisation activities relevant to COVID-19 in Spain and Italy illustrate – although fragmented and partially complete – the importance of standardisation in key sectors for combatting pandemics, such as in health, social, safety and security.

The outcome of the in-house scoping exercise highlights where standardisation can potentially help research and innovation to be more policy-goal-driven and to better support Europe's contribution to the global efforts in managing the pandemic through increased alignment and harmonisation.

This report informs colleagues in European institutions and Member States about the crucial role standardisation plays in the common efforts to overcome the COVID-19 pandemic. Examples include potential inputs to the drafting of guidelines, methods and or interoperability standards.

Finally, the report also provides practical examples of agile standardisation activities and deliverables that have the potential to enable the EU to respond more effectively and multilaterally to future crises. With this report we aim to raise awareness about the opportunities that standardisation and harmonisation can bring in the context of the COVID-19 pandemic.

# 1 Introduction

On 22 January 2020, while World leaders celebrated the 50th World Economic Forum in Davos to share positive economic projections, only one single briefing session was dedicated to Coronavirus[1]. By that time, the German Robert Koch Institute issued a statement[2] that assessed coronavirus as less dangerous than SARS, although the European Commission had set-up its coronavirus specific *Early Warning and Response System*, and the Health Security Committee had just concluded its first meeting[3]. Two months later, following the surge of COVID-19 cases, most countries in Europe adopted national and regional travel restrictions, isolation of outbreak areas and physical distancing measures to control and contain the spreading of the virus. A year later 125 million people had been reported infected and more than 2.8 million people had died from COVID-19 all over the world, from which 25 million infected, and more than 600,000 dead across Europe.

In order to contain the pandemic, Member States followed different responses and strategies, with very little harmonisation amongst each other. The European Commission, bound by the European Union subsidiarity principle, has no legal power to enforce health management policies or actions over Members States, such as quarantine measures or school closures. Nevertheless, the European Commission has joined forces with Members States for a coordinated response, such as preparedness actions (PREPARE[4]) and purchasing medical equipment including vaccines through a joint procurement mechanism.

After six months confronting the COVID-19 pandemic, it has become clear that several actions of Europe's COVID-19 response strategy could benefit from a more harmonised approach and standardisation activities. Travelling bans have been lifted and replaced with self-monitoring and public quarantine advices, centralised reporting has been setup, mask wearing was harmonised and EU-wide standardised protective equipment was produced for the EU market.

The European Union Single Market, as one of the great successes of the European Union, would not have been effective or even possible without standards: in addition to industry specifications on how to manufacture safe and secure goods, standards also play a key role in measuring and services, as well as in protecting the environment and ensuring the safety and security of people. Standards can therefore offer an effective vehicle to confront more effectively the COVID-19 pandemic through the adoption of harmonised approaches building up from consensus across all countries in Europe.

The COVID-19 response strategy of the European Commission encompasses several main elements[5]:

- *Economic measures:* recovery plan, pandemic emergency purchase programme, state aid, flexibility of the European fiscal framework, screening of foreign direct investment, EU Solidarity Fund, Providing economic guidance (European Semester), SURE[6] instrument protects jobs, SME liquidity measures, investment initiative.

- *Research & Vaccine strategy:* financial support to vaccine developers securing the production of vaccines in the EU, adapting the EU's regulatory framework to the current urgency.

- *Public health:* stockpile of equipment, medical guidance for Member States, panel of 7 independent epidemiologists and virologists, ECDC[7] community measures (i.e. physical distancing), guidelines and 'Clearing house for medical equipment', guidelines on testing methodologies, tracing mobile apps, Remdesivir (a pharmaceutical drug) marketing authorisation.

- *Personal protective equipment:* ramping up production, recommendation on conformity assessment, convert production lines, European standards for medical supplies, requirements for export authorisation, joint procurements.

- *Borders and mobility:* Border management measures, non-essential travel restrictions to the EU, free movement of workers, guidance to Member States (MS) on health workers, EU guidelines on green lane, repatriation of EU citizens, support to airlines.

---

[1] https://www.weforum.org/events/world-economic-forum-annual-meeting-2020/programme
[2] https://www.zdf.de/nachrichten/zdf-morgenmagazin/wieler-zu-coronavirus-gefahr-gering-100.html
[3] https://ec.europa.eu/info/live-work-travel-eu/coronavirus-response/timeline-eu-action_en
[4] Platform for European Preparedness Against (Re-)emerging Epidemics
[5] https://ec.europa.eu/info/live-work-travel-eu/health/coronavirus-response/overview-commissions-response_en
[6] European Instrument for Temporary Support to mitigate Unemployment Risks in an Emergency
[7] European Centre for Disease and Prevention Control

- *Fighting disinformation:* promotion of authoritative content, identification of 300 disinformation narratives, Rapid Alert System.

So how did the European Union coordinate its response to the COVID-19 pandemic and which technical fields require multilateral coordination?

In May 2020 the European Union and the United Nations published the Global Response Plans, providing a comprehensive and coordinated overview of the available knowledge regarding resilience and preparedness across Europe and beyond.[8]

Since the beginning of the crisis, the JRC has mobilised its diverse expertise in a dedicated in-house Task Force for COVID-19 to support the European Commission in understanding the characteristics of the emergency, anticipate its impacts and support contingency planning. This support to the Commission services was direct, providing expert advice, as well as through crisis coordination mechanisms, data and analysis.

These efforts required essential agreements on terminologies, methodologies, procedures and data formats. JRC scientists worked in the preparation of guidelines and control materials for testing and better characterization of the pandemic across different sectors (agriculture, energy supply, transport, etc.), but also transversally to evaluate and anticipate the evolution of the crisis and to effectively manage it.

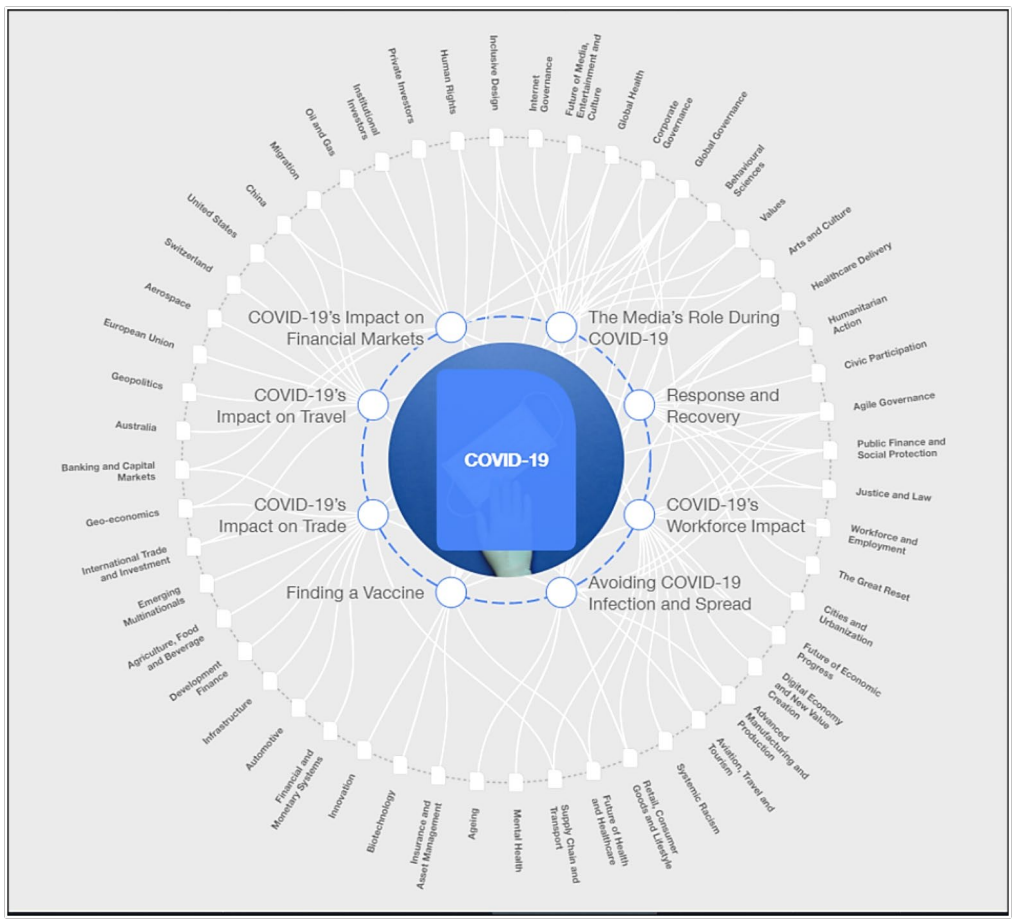

**Figure 1.** Current knowledge resources of eight COVID-19 issues and their association with common topics
Source: World Economic Forum, COVID-19 Strategic Intelligence, 2020.  https://intelligence.weforum.org/[9]

---

[8] https://ec.europa.eu/info/research-and-innovation/research-area/health-research-and-innovation/coronavirus-research-and-innovation_en
[9] https://intelligence.weforum.org/topics/a1G0X000006O6EHUA0?tab=publications; the Strategic Intelligence tool by the World Economic Forum provides an expedient platform to locate publications to specific Covid-19 related issues. The tool pictures over 50 areas of interest, from Arts and Culture to Geo-economics, and binds them with eight-mayor research questions related to the pandemic. By clicking on one research question, i.e. Finding a Vaccine, the tool points out the related areas of interest. Moreover, it shows scientific papers, published on this topic.

The tool published by the World Economic Forum (Figure 1) highlights the interconnectedness and multidisciplinary character of global response strategies and related research fields. Such intelligence solutions can help in providing rapid access to scientific information in high-specialised research fields. In order to support a firmer harmonisation of measurements, methodologies and rendered services, the first attempts of standardisation should be timely and efficient to avoid unnecessary efforts and time spent on the development of parallel solutions.

The JRC is well positioned in dealing with technical research tasks of multidisciplinary character that require multilateral coordination. Its Corona Tasks Force took over the coordination effort to interconnect all corresponding scientific research units at the JRC with EC policy services under one umbrella.

Therefore, the JRC in collaboration with CEN and CENELEC[10] launched in May 2020 an in-house scoping exercise on standardisation needs in response to COVID-19 and future pandemics. The purpose of this scoping exercise was to identify examples of both opportunities and standardisation needs in relevant sectors in consultation with different stakeholders, researchers and standardisation experts. This report reflects the results of this scoping exercise and provides practical examples of standardisation agile activities and deliverables that can help Europe coping with the present and future pandemics.

It serves predominantly to inform European public servants and the standardisation communities. The report presents the needs for standardisation that were identified for handling health, social, safety and security issues in a better and more efficient way during and after pandemics.

This report does not reflect the complete list of standardisation activities already ongoing on COVID-19 but mainly aims to provide meaningful examples in which standardisation can help research and innovation to be policy-goal-driven and help Europe to manage the pandemic with strong alignment and harmonisation.

---

[10] the European Committee for Standardization (CEN), the European Committee for Electrotechnical Standardization (CENELEC)

## 2  Where have standards been applied during the COVID-19 pandemic?

Standardisation makes use of international expertise and knowledge available to ensure that products and services meet quality, safety and performance requirements. A standard is a document that sets out requirements for a specific item, material, component, system or service, or describes in detail a particular method or procedure. As detailed below, existing standards can help in both mitigating the negative effects of the global COVID-19 pandemic and in developing recovery measures. The European Standardisation System is available to respond with new standardisation deliverables to address needs arising in both fighting the spread of the virus as well as in supporting post-pandemic recovery measures.

Examples of areas in which existing standards can be relevant:

*Medical devices and protective equipment*

Standards in areas such as medical gloves, medical facemasks, personal eye protection, etc. are particularly important to ensure that the equipment needed to confront COVID-19 meets requirements and is reliable. In March 2020, upon the urgent request from the European Commission (DG GROW), CEN and CENELEC in collaboration with their members have made available a series of European standards (ENs) for medical devices and personal protective equipment used in the context of the COVID-19 pandemic. The list of the 11 European standards EN(s) made available is retrievable[11].

An example of prompt standardisation response to the COVID-19 pandemic is the CWA[12] 17553:2020 deliverable 'Community face coverings – Guide to minimum requirements, methods of testing and use', which provides minimum requirements for the design, production and performance assessment of community face coverings (barrier masks) intended for consumers, single or reusable. This CWA has been developed in three weeks upon the urgent request from the European Commission and it is now available for free download from the CEN and CENELEC website and from the websites of CEN national members.

*Laboratory standards*

Standards ensure that medical tests provide accurate and reliable results. Laboratory standards provide high quality, high volume testing capabilities leading to comparable results and allow responding effectively to the COVID-19 pandemic. For COVID-19 testing laboratories it is critical to implement quality assurance through accreditation. Accreditation for COVID-19 testing in accordance with ISO 15189 (medical laboratories) and 17025 (general testing requirements) demonstrates competence to perform testing and guarantees accurate and reliable test results. Centres that manage bio risks, such as tests, storage, transport, working with or disposing hazardous biological materials employ ISO 35001 to safeguard their staff.

The Commission document C(2020)8037 recommends the use of rapid antigen tests for the diagnosis of SARS-CoV-2 infection and has developed test performance criteria that aim to improve the overall performance of these tests. This can benefit all European citizens and is key for of the exit strategy from the current crises (2020/C 122 I/01).

*Environmental and waste management standards*

These standards help to manage the increased waste caused by the use of single-use protective equipment (protective clothes, single-use gloves, medical masks, etc.). Besides occupational health and safety management, also environmental and waste management aspects are captured in the standard (ISO/PAS 45005:2020(en)).

*Security and resilience standards*

Business continuity management and emergency management is particularly important in the short-term but remains also a priority in the medium and long-term. Business continuity management serves as a guide for the design, implementation and maintenance of risk management. To ensure the operability, and therefore the survival of a company or government agency, suitable preventive measures must be taken to increase the robustness and reliability of the business processes as well as to enable a quick and targeted reaction in case of an emergency or a crisis.[13]

---

[11] https://ec.europa.eu/commission/presscorner/detail/en/ip_20_502
[12] CEN Workshop Agreement
[13] Crisis management PD CEN/TS 17091:2018; Business continuity management systems BS EN ISO 22313:2020; More information from British Standards Institute, NEN, and CEN-CENELEC and an example BSI 100-4 or UNIDO's COVID report

*Remote working and distant learning*

Under the COVID-19 pandemic, remote operation systems have driven one of the fastest changes to education across the globe. Circumstances dictated by social distancing measures have prevented partial or complete class room teaching. This has led to a switch to distance learning, which has required new practices for working and learning in remote conditions. Similarly, due to the COVID-19 pandemic, working from home has become the norm for millions of workers in the EU and worldwide. Early estimates from Eurofound (2020)[14] suggest that close to 40% of the workforce began to telework fulltime as a result of the pandemic. A recent JRC study[15] estimates that 25% of employees work in telework-able sectors. Considering that before the outbreak just 15% of those employed in the EU had ever teleworked, many people face currently a sudden shift to remote working. As a result, standards on ICT for remote working are being revisited.[16]

*Assisting tourism businesses*

Safe service to both consumers and workers assist the tourism businesses with solutions to maintain operations. Some CEN and CENELEC Members are developing guidelines to ensure safety solutions for the tourism industry following the COVID-19 emergency.

*E-health / ICT*

The Rolling Plan on ICT standardisation[17] compiles the European Commission's views and policies, views of other stakeholders and information on existing ICT standards work. The consultation of the different CEN Technical Committees (TCs) and CEN and CENELEC members will be organized under the responsibility of working group 6 of the technical board 'ICT standardization policy' for the next revision cycle.

In the context of the COVID-19 pandemic, the task force of the European Commission's Rolling Plan on ICT standardisation has decided recently to draft an addendum to the 2020 Plan to address the role that ICT standardisation can play, which includes proposed actions in support of:

- activities against SARS-CoV-2 and to protect citizens;[18]
- being better prepared for upcoming waves of COVID-19 or future pandemics;
- the recovery of European industry.

---

[14] Eurofound (2020), Living, working and COVID-19, COVID-19 series, Publications Office of the European Union, Luxembourg.
[15] https://ec.europa.eu/jrc/sites/jrcsh/files/jrc120945_policy_brief_-_covid_and_telework_final.pdf
[16] See BSI
[17] https://ec.europa.eu/digital-single-market/en/rolling-plan-ict-standardisation
[18] SARS-CoV-2 and its mutants is the virus that causes the COVID-19 pandemic. COVID-19 is the corona virus crisis.

# 3   Ongoing COVID-19 specific related standardisation initiatives by National Standardisation Bodies – examples of Spain and Italy

National standardisation bodies across Europe have been assisting national governments to facilitate a rapid response to COVID-19. National standardisation bodies are key in striving towards solutions that allow a controlled and secure economic recovery. In this sense, the European industry, and small and medium enterprises (SMEs) work on standards and protocols that can define the basis for allowing sector by sector a reopening of economic activities respecting minimum agreed standards.

A non-exhaustive list of initiatives undertaken by Spain and Italy is presented below:

- The Spanish Association for Standardisation (UNE), in collaboration with the Institute for Spanish Tourism Quality (ICTE), has published a series of UNE Specifications, which establish the guidelines and recommendations for reducing the risk of the spread of coronavirus SARS-CoV-2 in the tourism sector.[19] UNE has started to develop specifications on the following topics:
    - Smart Cities: City platform for public health services in emergency situations;
    - e-Health: medical teleconsultations;
    - Guidelines for working at home;
    - Protocol for reactivation of fairs and congresses under COVID-19 risk;
    - Specification for unloading fish at ports;
    - Small retail businesses;
    - Recreational diving services;
    - Guidelines for safe cinematographic filming sets;
    - Guidelines for safe passenger transport;
    - Protocol for carrying out tests without leaving the vehicle;
    - Use of urban parking for COVID-19 tests: Guide to identify and evaluate parking lots for the carrying out COVID-19 tests.

- The Italian Standardisation Body (UNI) is developing the following specifications:
    - Distance learning in schools: Operational guidelines for planning distance learning in schools. The document is addressed to schools and aims to deal with the current sanitary emergency and in Implementing new innovative learning methodologies (for the future);
    - Safety in the tourism industry following COVID-19: Guidelines to ensure safety solutions for the tourism industry following the COVID-19 emergency. It aims to assist tourism businesses in providing a safe service to both consumers and workers. It will be a framework document in which common solutions for everyone and specific provisions for the different sectors are described;
    - Safety signs: Guidelines on safety signs for COVID-19.

Other national standardisation bodies updated their work programmes as well, with many standardisation activities and updates in response to the COVID-19 pandemic. These updates are for example well visible at the British Standards Institute and Royal Netherlands Standardization Institute.

---

[19] https://www.en.une.org/la-asociacion/sala-de-informacion-une/noticias/directrices-para-un-turismo-seguro

# 4 Opportunities and key standardisation needs identified by the JRC

## 4.1 How the scoping study was conducted?

A scoping exercise on standardisation needs that supports an effective response to COVID-19 and future pandemics was carried out within the JRC starting in March 2020. The purpose of this scoping exercise was to identify opportunities and standardisation needs in relevant sectors in consultation with different stakeholders, researchers and standardisation experts.

A rapid questionnaire was circulated in March 2020 to the JRC staff that was registered within the JRC Corona Task Force. Based on this questionnaire, which included proposal and a series of teleconferences between CEN CENELEC and members from the JRC Corona Task Force, needs were identified by the JRC scientists that responded to the call, describing technical areas were harmonisation efforts would yield potential benefits in managing the COVID-19 pandemic.

A form was circulated via email and the JRC intranet, requesting an overview of existing standards, methodologies, procedures, and guidelines relevant in confronting present and future pandemics. In a second part, information about potential relevant standardisation needs were requested.

The following topics were addressed:

- Methods to share epidemiological information;
- Improved crisis management and business continuity capabilities through standards;
- Method to compute the Effective Reproduction number from data series;
- Guidance for quality assurance in testing and detection laboratories.

These topics were assessed by the JRC and brought to the attention for review by CEN-CENELEC. This led to a series of meetings and discussions between the JRC scientists that proposed these topics and experts involved in CEN and CENELEC Technical Committees. A potential incorporation of those topics in the work of technical working groups is being explored.

The following sub-sections provide a brief overview of the four topics listed above.

## 4.2 Methods to share epidemiological information

The EU's common efforts in confronting the virus are hampered, among others, by the current fragmentation of data sharing. Several methods exist to share epidemiological data across public institutions. However, few of them are interoperable and often the quantities proposed are different or characterised with different taxonomy. Aggregated epidemiological indicators (e.g. number of cases, deaths, hospitalised cases…) shared by Member States could be improved by format standardisation, for example when provided by Member States through various types of web applications (e.g. webpage, web service, dashboard, web map application).

The standardised data format of aggregated indicators outlined in this proposal does not refer to existing reporting through systems such as The European Surveillance System (TESSy), but to web applications and dashboards developed voluntarily by Member States and not part of reporting requirements according to EC 851/2004 and Decision 1082/2013/EU.

While the level of aggregation of the data should be as detailed as possible and as a minimum at regional level[20], a minimum set of common, comparable indicators for key epidemiological data across Member States could include:

- Cumulative number of positive cases;
- Cumulative number of fatalities;
- Cumulative number of recovered cases;

---
[20] Following the Nomenclature of Territorial Units for Statistics (NUTS2), dated 2016, including 244 regions at NUTS 2 or 1215 regions at NUTS 3 level,

- Hospitalised cases (both current number and new admissions of COVID-19 confirmed cases);
- COVID-19 confirmed cases in intensive care units (both current number and new admission of COVID-19 confirmed cases);
- COVID-19 tests performed;
- ISO-3 country code;
- Aggregation according to the nomenclature of territorial units for statistics (NUTS2 or NUTS3) of subnational data;

CEN/TC 251 "Health informatics" covers standardisation in the field of Health Information and Communications Technology (ICT) to achieve compatibility and interoperability between independent systems and to enable modularity. This includes requirements on health information structures to support clinical and administrative procedures, technical methods to support interoperable systems as well as requirements regarding safety, security and quality. More specifically, the scope of its WG 2 "Technology and Applications" includes also standardisation of data formats used for the exchange of information from medical and other health devices to other systems.

CEN Technical Committee 251 is a technical decision making body working on standardization in the field of Health Information and Communications Technology in the European Union. The goal is to achieve compatibility and interoperability between independent systems and to enable modularity in Electronic Health Record systems.

Examples of relevant published standards include CEN/TR 15253:2005 Health informatics – Quality of service requirements for health information interchange, and EN ISO 21549-1:2013 Health informatics and patient healthcare data (general structure, common objects, limited clinical data, extended clinical data, identification data, administrative data, medication data, links).

At international (ISO) level, this area is covered by ISO/TC 215 "Health informatics", which addresses standardisation in the field of health informatics to facilitate capture, interchange and use of health-related data, information, and knowledge to support and enable all aspects of the health system.

Possible stakeholders in standards process to identify methods to share epidemiological information could include:

- Joint Research Centre of the European Commission;
- Health and Food Safety (SANTE) Directorate-General of the European Commission;
- European Centre for Disease Prevention and Control (ECDC, EU Agency);
- Health Security Committee (HSC);
- World Health Organisation's Regional Office for Europe (WHO EURO).

The standardisation of the method and format to share COVID-19 epidemiological indicators should apply to a minimum set of common, comparable indicators of key epidemiological data. This data are voluntarily provided by web applications and dashboards developed at national and sub-national level.

The current fragmentation of approaches for data sharing, for instance some aggregated epidemiological indicators (number of cases, deaths, hospitalised cases…) shared by Member States, are causing lack of interoperability and hence the shared data is not available for global comparisons.

To fully unlock the opportunities for modelling and carrying out timely analysis, there is a need to work on a common and comparable simple list of indicators for key epidemiological data (see below) across Member States at the same geographical scale.

In particular, to effectively model the expected epidemic trend at European level, the collection, harmonization and sharing of epidemiological data from national websites at subnational level should be encouraged.

The topic of sharing epidemiological data is of great interest to CEN and CENELEC, because of the clear European context of the topic and because it possesses an in-house capacity that can ensure broad and diverse stakeholder engagement from their national members and networks.

CEN Technical Committee 251 is considering a possible inclusion of standardisation needs from the topic of sharing epidemiological data in its work programme for 2021. TC 251 escalated this request to its Joint Initiative Council on COVID-19. Hereafter the brief summary of the enquiry to its members:

The suggested reporting format has a different scope than regulated data, hence does not refer directly to existing reporting through systems such as TESSy [the European Surveillance System] from ECDC. It aims to support web applications and dashboards developed voluntarily by Member States and is not part of formal reporting requirements. At the moment several methods exist to offer data for public and institutional evaluations (text files, xml files, json, web pages, images, dashboards). None of them is interoperable and often the quantities proposed are different or with a different taxonomy. The members of the Joint Initiative Council have been asked whether they are aware of existing standards or initiatives for voluntary public reporting of COVID-19 epidemiological data that can be (re)used.

The database and universal standard for identifying medical laboratory observations, the Logical Observation Identifiers Names and Codes (LOINC), a part of the COVID-19 Interoperability Alliance, is in the process of creating standardised terminology related to SARS-CoV-2 laboratory tests, including test results reporting and COVID-19 case reporting, as well as clinical notes. LOINC established 3 groups for SARS-CoV-2 laboratory terms, as well as a Fast Healthcare Interoperability Resource (FHIR)[21] value set for terms related to case reporting.

In a joint virtual conference held on 2 October 2020 that included the participation of CEN TC 251, JRC and ECDC, the following steps towards assessing standardisation needs were agreed:

- JRC, DG SANTE and ECDC to further clarify the scope of a potential standardization activity;
- ECDC to discuss with its national network to check the appetite for this initiative;
- CEN TEC 251 and JRC to identify a list of "pilot countries" to discuss with.

## 4.3 Improved crisis management and business continuity capabilities through standards

The current pandemic has shown that disasters do not stop on borders. There is a need to mitigate risks, in particular those with a cross-border impact, by improving the coordination and capabilities among Member States. This need may be broken down into the following aspects:

- *Mapping of capabilities* required to mitigate risks, i.e. performance tests of emergency equipment such as personal protective equipment (PPE) and medical equipment, tests required for general data protection for deterring chemical, biological, radiological, nuclear, and explosives threats related to the security technology regulation (GDPR), and detection technologies for deterring chemical, biological, radiological, nuclear, and explosives threats (CBRNE) with ensured interoperability and shared understanding of the security and emergency terminology;
- *Coordination* mechanism among EU Member States, reinforced by guidelines on closing borders, closing shops, etc.;
- Developing a crisis management capability through *preparedness*, i.e. business continuity, education and training of staff on crisis/disaster management, communication, alternative methods for efficient continuity of activities;
- *Guidance* documents regarding communication to the public, i.e. guidelines on digital environment and on risks, control, damage limitation and fake news.

Most aspects can be reasonably addressed in a long-term strategy with future policies to improve the preparation of the society for possible future pandemic crises; a standardisation work programme on societal resilience should be part of this vision. This could entail updating existing standards, as well as developing new standards to meet emerging needs.

---

[21] FHIR is a standard describing data formats and elements (known as "resources") and an application programming interface (API) for exchanging electronic health records (EHR).

Several EU policy statements stress the importance of a harmonised disaster preparedness and improved resilience in Europe.[22] Moreover, Europe maintains a disaster preparedness plan (the 'rescEU reserve') which includes a pre-identified fleet of firefighting planes and helicopters, medical evacuation planes, as well as stocks of medical equipment and field hospitals that can respond to health emergencies, and chemical, biological, radiological, and nuclear incidents.

Standardisation can help providing agreed specifications, technical guidance and tools to deal with current and future pandemics. European standards can ensure that the same specifications are adopted in all Member States, supporting the coordination and alignment of national responses.

Standards are needed to keep people safe, to keep results reliable and comparable, and to build confidence in the process of managing disasters such as the COVID-19 pandemic. There is a critical need for laboratories, especially in countries that are lacking particular resources and capabilities to establish clear guidelines when handling biological agents. In many countries, especially those without any national policies or regulations in place, users would benefit from standards.

A strong, resilient, green, just and inclusive recovery from the COVID-19 pandemic can be achieved when consumers, enterprises and regulators collaborate and agree, for example, on standards that advise on working safely in a pandemic (ISO/PAS 45005), that guide on how to manage biorisk (ISO 35001), and on how to manufacture facemasks (EN 14683). It is also important that standards address regulations such as the conduct of clinical trials (EU 2020/1043), recommendations on a common approach to the restriction of free movement (EU 2020/1632) and common testing strategies, including the use of rapid antigen tests (EU 2020/1595).

Standards build trust in products and services and act as an effective tool for coordination and solidarity to fight crisis situations.

Standardisation activities in the field of societal resilience and disaster preparedness at the European level relate to security platforms addressed by the following:

- CEN and CENELEC Coordination Group on security;
- CEN/TC 391- Societal and citizen security;
- CEN/TC 439- Private security services;
- CEN Strategic Advisory Body on Environment (SABE)
- CEN-CENELEC Sector Forum PPE;
- CEN-CENELEC Sector Forum Medical devices;
- CEN/ CLC/ JTC 13 'Cybersecurity and data protection';
- ISO/TC 292 "Security and resilience ;
- ISO/TC 262 "Risk management".

CEN/TC 391 would be the most logical host for the development of the standardisation work programme (e.g. CEN/TC 391/WG 3, Crisis management/civil protection). Therefore, the initiative of the JRC in bringing the topic discussed on this section to CEN/TC 391 is beneficial and could encourage a joined effort by EU Member States.

Potential and not exclusive stakeholders in a standardisation processes addressing societal resilience are:

- Security industry (detection technologies, private security providers, emergency services);
- Emergency services;
- PPE (Personal Protective Equipment) industry;
- Producers of medical devices;

---

[22] COM(2017) 610 final Action Plan to enhance preparedness against chemical, biological, radiological and nuclear security risks; COM(2019) 640 final The European Green Deal

- Medical laboratories;
- Certification bodies;
- Horizon 2020 projects funded under Work Programme 2018-2020 Action 14: Secure societies-Protecting freedom and security of Europe and its citizens" calls addressing, among others, disaster resilient societies, technology for first responders, CBRN and novel concepts for the management of pandemic crisis.

Stakeholder engagement and mobilisation is crucial to advance activities on societal resilience, such as organising a Focus Group at CEN.

## 4.4 Method to compute the Effective Reproduction number from data series

Public authorities in charge of monitoring the spread of COVID-19 are in need of defining the progress of the pandemic growth. After the introduction of containment measures, it is crucial to provide a method to show the efficiency of the implemented measures. During the following phase of de-escalation there is a great interest in understanding the need to reintroduce containment measures, based on the probability of a new outbreak. A vital indicator is the Effective Reproduction number ($R_t$), which represents the number of secondary infections caused by every single infected person. $R$ represents the expected number of cases directly generated by one case in a population where all individuals are susceptible to infection.[23]

In practice the time-dependent effective reproductive number $R_t$ [24] can be estimated from an epidemiological curve based on the date of onset or through more advanced statistical and mathematical models. It is useful for longitudinal follow-up in transmission trend.

The value of $R_t$ can be calculated with different methods, starting from the same data series of infected or affected population. At the moment, almost every country in Europe adopts its own standard method for calculating this value and little harmonisation exists including follow up decisions.

Benefits of global pandemic influenza surveillance affects every country. Sharing standardised and coordinated international information is crucial for crisis management at global and national levels. National authorities need to know how the pandemic is evolving not only in their own country, but also in neighbouring countries and continents. Sharing the information at global level is beneficial to all.

Although many mathematical models have been developed to estimate several types of reproduction numbers during epidemic outbreaks, there is no unique work frame. More than 615,000 publications on the "Effective Reproduction number ($R_t$)" have been published during the past 50 years, of which 21,300 were published within 2019 and 2020.

There is still no standard method to define the number, but several methods are available on open source platforms such as github. Several reports are comparing and evaluating these methods.

At international level, there are many standardisation activates on statistics and data interpretations.

Relevant International Standards, Manuals and Guidelines:

- ISO 3534-1:2006 Statistics — Vocabulary and symbols (TC69) Standardisation in the application of statistical methods, including generation, collection (planning and design), analysis, presentation and interpretation of data: General statistical terms and terms used in probability, applied statistics, design of experiments, survey sampling.
- ISO 16269-6:2013 Statistical interpretation of data — Part 6: Determination of statistical tolerance intervals. Standardisation in the application of statistical methods, including generation, collection (planning and design), analysis, presentation and interpretation of data.

---

[23] $R = \tau\ c\ d$ Where $\tau$ is the probability of transmission, which can be influenced by the use of protective equipment, c is the contact rate among individuals and d is the duration of the infectious period. The basic reproduction number ($R_0$) measures the transmission potential of a disease. It is the average number of secondary infections produced by one case in a population where everyone is susceptible to infection. In practice, for new emerging pathogens, $R_0$ is often estimated from the initial growth rate of the outbreak (i.e. derived from the epidemiological curve, or contact tracing data).

[24] When population immunity is high enough to reject the hypothesis of full susceptibility, or when response measures are in place, then a second indicator is often used: *the time-dependent effective reproductive number* ($R_t$, sometimes $R_e$ or $R_{eff}$).

A selection of relevant European manuals, guidelines and WHO references is provided in Annex 2.

Eurostat and national statistical authorities have been working together to elaborate guidelines and notes on how to address the methodological issues triggered by COVID-19 in data statistical production. Their intention is to ensure that European statistics continue to be based on sound foundations.

The following relevant stakeholder groups have been identified:

- World Health Organization (WHO);
- Governments and National authorities;
- Relevant European Commission services including the JRC;
- Statistical office of the European Union (Eurostat);
- European Centre for Disease Prevention and Control (ECDC);
- International Statistical Institute (ISI);
- Research facilities/ control laboratories;
- National public health organizations;
- Health care system;
- Data management industries.

To conclude, there is no international or European standard method for computing the Effective Reproduction number from data series. Moreover, there is an urgent need to define such standard. The Effective Reproduction number is a key parameter for tracing the evolution of an infectious disease pathogen like COVID-19, i.e. the expected number of secondary cases per primary case in a completely susceptible population. A harmonised method will help health care systems, authorities, governments and researchers to have reliable and accurate statistics that can be shared and easily compared among different actors.

Next steps include bringing the topic to the attention of CEN-CENELEC Members to identify relevant TCs. The JRC could address the topic and identify, together with CEN-CENELEC, relevant platforms. Further, Eurostat, as well as ECDC and ISI should be involved.

## 4.5 Guidance of Quality Assurance in testing and detection laboratories

SARS-CoV-2 testing laboratories have been challenged with many issues during the evolution of the COVID-19 pandemic. Assuring quality was one of such challenges. Large variability in PCR test results may have been caused in part to a not full comprehension of the 'window period' after virus acquisition in which testing is most likely to produce false-negative results (sampling issues).[25] The sensitivity and specificity of PCR-based tests for SARS-CoV-2 are often inadequately characterised. Needs identified in the EU include support with sequencing capacities and protocols and with bioinformatics in particular.[26] Although EU countries stepped up dramatically their testing capacities over the past year, shortages of laboratory consumables and human resources, as well as sample storing facilities, continue to exist and could affect the overall laboratory response to COVID-19. Both ISO/IEC 17025:2017 and ISO 15189:2012 address issues of sampling, method validation and other processes and resource requirements that may impact the reliability of a result. Promoting these standards and making them more accessible in the form of e-courses is one way the EU Academy[27] and JRC aim to ameliorate the deficiencies.[28]

The capacity for the identification of SARS-CoV-2 variants using sequencing technologies is according to an ECDC assessment in several EU Member States below the recommendation set by the European Commission. Not all of the EU Member States are investigating the emergence of SARS-CoV-2 variants, but are increasing or planning to increase their sequencing capacity. In a recent assessment Member States indicated the need for support from ECDC.[25]

---

[25] ECDC rapid assessment of laboratory practices and needs related to COVID-19
[26] Detection and characterisation capability and capacity for SARS-CoV-2 variants within the EU/EEA
[27] The EU Academy is an EU-owned online hub containing first-hand knowledge, high quality educational resources and valuable insights, directly produced by the EU institutions, for individuals whose work is related to its sphere of action
[28] Communication from the Commission Guidelines on COVID-19 in vitro diagnostic tests and their performance 2020/C 122 I/01

*(Bio)-Chemical Testing*

Relevant international standards for (bio)-chemical testing are:

- ISO/IEC 17025:2017 General requirements for the competence of testing and calibration laboratories. This standard has been developed with the objective of promoting confidence in the operation of laboratories. It contains requirements for laboratories to enable them to demonstrate they operate competently and can generate valid results. Laboratories that conform to this standard will also operate generally in accordance with the principles of ISO 9001.

- ISO 15189:2012 Medical laboratories — Requirements for quality and competence. This standard can be used by medical laboratories in developing their quality management systems and assessing their own competence. It can also be used for confirming or recognising the competence of medical laboratories by laboratory customers, regulating authorities and accreditation bodies.

ISO IEC 17025:2017 enables laboratories to demonstrate that they operate competently and generate valid and reliable results, thereby promoting confidence in their work both nationally and around the world. European Regulations and directives require that measurements performed in accredited laboratories fulfil the requirements of this standard (as well as ISO 15189). Detection laboratories involved in COVID-19 measurements need to demonstrate their competence as well.

The JRC is currently preparing a guidance for laboratories carrying out (bio)-chemically based testing in the form of an online course entitled 'Quality Assurance for testing laboratories' and will lean on the ISO 17025 standard for its main components with examples and assignments being elaborated with the relevant target audience (including COVID-19 testing laboratories). The e-course will be available at the EU Academy in mid-2021.

*Biomedical testing*

Diagnostic tests should provide accurate and reliable results enabling to gather data from various laboratories and to allow comparison between region/countries. Several parameters characterising the different tests available on the market should be harmonised, assessed in similar conditions with reference materials, with clear procedures and with a clear description of the expected results.

Tests must be sensitive with a clear definition of their LOD (limit of detection) and LOQ (limit of quantification). LOD and LOQ describe the smallest concentration that can be reliably measured by an analytical procedure. The specificity of a test describes to what extent closely related non-target organisms cross-react with target agents. Reproducibility refers to the inter-assay variance, the variation of the results between runs or between laboratories. Repeatability refers to intra-assay variance which is the robustness of the assay, the variation of the results with the same sample analysed repeatedly in the same assay.

The minimum parameters to be included in a statistical analysis for performance analysis are the sensitivity ($S_e$), the specificity ($S_p$), the positive predictive value (PPV) and the negative predictive value (NPV). The number of true positives (PPV) and negatives (NPV), and false positives and negatives for the respective agents should be calculated and used to calculate the upper and lower confidence interval (CI), usually at 95%.

The setting of reliable testing procedures requires a reference material (the agent or a simulant) and reliable sample preparation methods. Clear guidelines for testing and assessment of functional characteristics are required. Comparison with a reference identification technology increases the confidence in the testing.

Besides analytical parameters, operational parameters may also be available in a standardised way such as the number of samples that can be processed simultaneously, sample preparation, sample volume, run time, time-limit for setting the analysis and for reinitiating the analysis , automated detection mode, data format, etc.

Adapting the format of providing guidance for SARS-CoV-2 (and other) testing laboratories via the creation of e-courses as well as the provision of a suitable reference material is one of the contributions of the JRC to respond to this crisis.

# 5 Conclusions

As COVID-19 emerged as a severe public health emergency for citizens, societies and economies in all Member States, a major economic shock has put Europe's well-being and development at stake. With the purpose of confronting the COVID-19 pandemic, the EU has prepared a global response plan to mitigate the socio-economic impact of the COVID-19 outbreak. The response plan shows clearly that an effective answer must be united, that cohesion is crucial and actions are harmonised and implemented recognising local conditions and needs.

The European Commission's role in enabling and coordinating the implementation is of vital importance. It builds on long-standing and successful experiences in harmonisation of interventions, procedures, and most importantly in emergency responses and crisis management.

Standardisation is an enabler that can deliver an efficient and effective response as part of the strategy adopted by the European Union to confront COVID-19. This report provides representative examples of the current COVID-19 related standardisation activities. It reflects on further needs as expressed by scientists at the JRC and standardisation organisations. The cases described in this report highlight possible scenarios on how standardisation organisations can bring them up in the corresponding fora.

This report aims at informing colleagues in European and Member State institutions about the crucial role standardisation plays in the common efforts to overcome the COVID-19 pandemic. Examples include inputs to a potential guideline, a method and an interoperability standard.

**Annexes**

**Annex 1. European Standardisation deliverables**

- **European Standard (EN)** is a normative document made available by CEN/CENELEC in the three official languages. The European Standard is announced at national level, published or endorsed as an identical national standard and every conflicting national standard is withdrawn.

- **Technical Specification (CEN/TS)**, is a normative document made available by CEN/CENELEC in at least one of the three official languages, that serves as normative document in areas where the actual state of the art is not yet sufficiently stable for a European Standard; The Technical Specification is announced and made available at national level, but conflicting national standards may continue to exist.

- **Technical Report (CEN/TR),** is an informative document made available by CEN/CENELEC in at least one of the official languages, established by a technical body and approved by simple majority vote of CEN/CENELEC national members. A Technical Report gives information on the technical content of standardisation work.

- **CEN Workshop Agreement (CWA),** which aims at bringing about consensual agreements based on deliberations of open Workshops with unrestricted direct representation of interested parties. This deliverable is particularly suitable to rapidly address new standardisation needs arising, as the average time to develop a CWA is 6-18 months.

**Annex 2. A non-exhaustive list of R references**

- *Global Surveillance during an Influenza Pandemic by WHO (2009)* Successful containment or control of pandemic influenza will rely on early recognition of sustained human-to-human transmission, which requires a system for outbreak detection, rapid data collection, analysis, assessment and timely reporting. An influenza pandemic will affect every country. Standardized and coordinated international information sharing is crucial for crisis management at global and national levels. National authorities will need to know how the pandemic is evolving not only in their own country, but also in neighbouring countries and continents.

- **Pandemic Influenza Risk Management by WHO (2017)** A WHO guide to inform and harmonize national and international pandemic preparation and respond

- *Estimation of $R_0$ and Real-Time Reproduction Number from Epidemics (RO Package)* by Pierre-Yves Boelle, Thomas Obadia: The *RO* package implements several documented methods. It is therefore possible to compare estimations according to the methods used. Depending on the methods requested by user, basic reproduction number (commonly denoted as *RO*) or real-time reproduction number (referred to as *R(t)*) is computed, along with a 95% Confidence Interval.

- *The $R_0$ package*: a toolbox to estimate reproduction numbers for epidemic outbreaks by BMC Medical Informatics and Decision Making (Year?): This work provides a review of generic methods used to estimate transmissibility parameters during outbreaks. Two categories of methods were available: those estimating the initial reproduction number, and those estimating a time dependent reproduction number. Five methods were implemented as an R library, developed sensitivity analysis tools for each method and provided numerical illustrations of their use. A comparison of the performance of the different methods on simulated datasets is reported.

- *Effective reproduction number estimation by Michael Höhle*: A licensed work under a Creative Commons Attribution-ShareAlike 4.0 International License. The markdown+Rknitr source code of this blog is available under a GNU General Public License (GPL v3) license from github.

*RWEEKLY.ORG*– Weekly Updates from the Entire R Community: R Weekly was founded on May 20, 2016. *R* is growing very quickly, and there are lots of great blogs, tutorials and other formats of resources coming out every day. *R* Weekly wants to keep track of these great things in the *R* community and make it more accessible to everyone.

## Annex 3. Guidelines and manuals on *R* with European focus

- Quality Assurance Framework of the European Statistical System: The ESS QAF represents a collection of methods, tools and good practices that are suggested for use and/or are already in use in (some of) the statistical authorities of the European Statistical System, where they have proved to be useful.

- European Statistical System handbook for quality and metadata reports: The ESS Handbook for Quality and Metadata Reports (EHQMR) is recognised as an ESS standard. It is included in the Catalogue of ESS standards, the collection of non-legislative normative documents underpinning the ESS. It is therefore a visible component of the ESS standardisation process, the importance of which goes well beyond making the current body of standards accessible.

- Towards a harmonised methodology for statistical indicators (Eurostat Manuals and Guidelines): Indicator typologies and terminologies, Communicating through indicators, Relevance for policy making

**GETTING IN TOUCH WITH THE EU**

**In person**

All over the European Union there are hundreds of Europe Direct information centres. You can find the address of the centre nearest you at: [https://europa.eu/european-union/contact_en](https://europa.eu/european-union/contact_en)

**On the phone or by email**

Europe Direct is a service that answers your questions about the European Union. You can contact this service:

- by freephone: 00 800 6 7 8 9 10 11 (certain operators may charge for these calls),

- at the following standard number: +32 22999696, or

- by electronic mail via: [https://europa.eu/european-union/contact_en](https://europa.eu/european-union/contact_en)

**FINDING INFORMATION ABOUT THE EU**

**Online**

Information about the European Union in all the official languages of the EU is available on the Europa website at: [https://europa.eu/european-union/index_en](https://europa.eu/european-union/index_en)

**EU publications**

You can download or order free and priced EU publications from EU Bookshop at: [https://publications.europa.eu/en/publications](https://publications.europa.eu/en/publications). Multiple copies of free publications may be obtained by contacting Europe Direct or your local information centre (see [https://europa.eu/european-union/contact_en](https://europa.eu/european-union/contact_en)).